"How trustworthy is this research?"

Designing a Tool to Help Readers Understand Evidence and Uncertainty in

Science Journalism

Anders Sundnes Løvlie
IT University of Copenhagen
asun@itu.dk

Astrid Waagstein
University of Copenhagen
awaa@hum.ku.dk

Peter Hyldgård
ScienceCom.dk
peter@sciencecom.dk

**Acknowledgments**

Numerous colleagues and students made valuable contributions to the work presented in this article. We wish to thank in particular the entire staff of videnskab.dk and the participating experts from the Nordic Cochrane Institute, as well as Martin Rust Priis Christensen, Louis Valman Høffding Dyrhauge, Søren Gollander-Jensen, Allan Linneberg, Anders Steen Mikkelsen and Jeppe Nicolaisen.

**Funding information**

This work was supported by the Google Digital News Innovation Fund.

# Abstract

Widespread concerns about the spread of misinformation have gained urgency during the ongoing COVID-19 pandemic and pose challenges for science journalism, in particular in the health area. This article reports on a Research through Design study exploring how to design a tool for helping readers of science journalism understand the strength and uncertainty of scientific evidence in news stories about health science, using both textual and visual information. A central aim has been to teach readers about criteria for assessing scientific evidence, in particular in order to help readers differentiate between science and pseudoscience. Working in a research-in-the-wild collaboration with a website for popular science, the study presents the design and evaluation of the Scientific Evidence Indicator, which uses metadata about scientific publications to present an assessment of evidence strength to the readers. Evaluations of the design demonstrate some success in helping readers recognize whether studies have undergone scientific peer review or not, but also point to challenges in facilitating a more in-depth understanding. Insights from the study point to a potential for developing similar tools aimed at journalists rather than directly at audiences.

# Keywords

Science journalism, design, human-computer interaction, health, evidence, uncertainty, misinformation, COVID-19

# Introduction

Widespread concerns about the spread of misinformation (Bechmann 2020; Lazer et al. 2018; McIntyre 2018) have gained urgency during the ongoing COVID-19 pandemic (Nielsen et al. 2020) and pose challenges for science journalism, in particular in the health area. This paper reports on a Research through Design (Zimmerman, Forlizzi, and Evenson 2007) project in collaboration with the Danish popular science website Videnskab.dk, exploring how to communicate the strength of the scientific evidence behind news items about medical research in a way that is accessible and understandable for a non-expert audience. The project brings together design and journalism research in order to address long-standing issues in science journalism concerning how to negotiate the sometimes conflicting concerns of scientists, journalists and audiences (Amend and Secko 2012; Weigold 2001).

Uncertainty is a key element in understanding science and the results that science produces. The way journalists work with and represent *epistemic* uncertainty, being uncertainty about the validity of truth claims, is thus critical to how the public understands and navigates in scientific knowledge presented to them by the media (Peters and Dunwoody 2016). Journalists and science communicators have developed different text based approaches to how and to what degree they deal with uncertainty in their work (Simmerling and Janich 2016). With the written word still being the basis of much journalism, less work has been done on how to use *visual* indicators to help audiences to decode/understand the epistemic uncertainty presented to them. Visual clues can take a key role in helping the users navigate in the many and complicated pieces of information they meet. This becomes even more crucial when discussions about science take place on social media, where many participants are reacting to headlines and visual 'clues' like pictures or videos, without reading the journalistic articles.

In this paper we explore how to design a tool for helping readers understand the strength and uncertainty of scientific evidence in news stories about health science, using both textual and visual information. The main objective is to teach the website's readers about criteria for assessing scientific evidence, in particular in order to help the readers differentiate between science and pseudoscience. The Scientific Evidence Indicator project makes a novel contribution to this area by addressing the need of science journalism to explain complicated scientific insights with appropriate rigor and accuracy, against the need to compete for the attention of the audience and offer short, simple and attractive presentations, assuming that the audience may only be spending very limited time and effort to read and understand the news stories. Thus, the study aims to strike a difficult balance between rigor and simplicity, to develop a tool that can help audiences without any specialist knowledge discriminate science from pseudo-science. Using a Research through Design (Zimmerman, Forlizzi, and Evenson 2007; Gaver 2012) approach, the study makes two types of knowledge contribution: First, through the presentation of an artifact (a tool) which addresses the challenges outlined in this article - which in the terminology of Human-Computer Interaction may be considered an 'artifact contribution' (Wobbrock and Kientz 2016); and second, the design process and evaluation offer some insights that may help us further understand the challenges and possible paths forward in communicating about scientific evidence to a general audience.

Research within health communication has demonstrated that audiences struggle to assess unreliable health claims in the media (Boutron et al. 2019; Haber et al. 2018; Walsh-Childers et al. 2018). This carries risks for individuals' decisions about their health, and may contribute to non-evidence-based practices which are widespread (Brownlee et al. 2017; Glasziou et al. 2017). In the context of the ongoing COVID-19 pandemic, concerns about misinformation have gained increased urgency due to controversies regarding vaccine hesitancy, contagion control measures such as facemasks and social distancing, and so on. These problems demonstrate a great need to educate the public to better understand how to distinguish between trustworthy and untrustworthy scientific claims regarding health.

## Science journalism and scientific uncertainty

Contemporary science journalism faces a dual challenge, consisting of two different problems: On the one hand, the need to identify and accurately report on *reliable* information about new developments in science, while avoiding to report unreliable information presented as (pseudo-)science. This problem is not new, but arguably the post-truth phenomenon and science denialism (McIntyre 2018; Diethelm and McKee 2009) have intensified this problem in recent years, making it ever harder to distinguish reliable information from misinformation, pseudoscience, commercial campaigns, political campaigns, conspiracy theories, fake news and so on (Weingart and Guenther 2016).

On the other hand, science journalism also needs to report openly and accurately on scientific *uncertainty*. This is a well-recognized challenge in research on science communication (Peters and Dunwoody 2016; Guenther and Ruhrmann 2016). Some studies have found that science journalism often fails to adequately communicate about scientific uncertainty (Matthias, Fleerackers, and Alperin 2020). Lehmkuhl and Peters (2016) find that the lack of attention given to scientific uncertainty is not primarily caused by the way journalism communicates, but rather is due to the journalists' perception of uncertainty (or lack thereof). Considering journalism about health research, Hinnant and Len-Ríos (2009) find that journalists struggle to balance a concern for scientific credibility with the need to simplify technical language to improve audience comprehension. Viviani and Pasi point to a lack of research on automatic credibility assessment of online health information, specifically in social media (Viviani and Pasi 2017). Turning to the audience, studies have found that communicating about uncertainty in science journalism is not necessarily detrimental for trust in science reporting (Retzbach and Maier 2015; Jensen 2008; Gustafson and Rice 2019). Takahashi and Tandoc (2016) suggest that a lack of trust in the news media may even work as an incitement to learn about science.

The scientific community puts much emphasis on gauging uncertainty. This is especially true for health science where e.g. the Hierarchy of Evidence is a generally accepted and widely used way of judging the importance of a scientific result on how doctors treat diseases and health care systems handle health related issues in society (Evans 2003; Guyatt et al. 2008). Some science journalists use such indicators on scientific uncertainty in their research/evaluation to judge whether to include a piece of scientific knowledge in their work. However, knowledge about these indicators is not widespread among journalists and the wider public. Instead journalists use other means to ensure that they can trust a given source, such as talking to other scientists, or using their own understanding of the scientific process and knowledge of scientific sources to assess the plausibility of claims (Peters and Dunwoody 2016; Guenther and Ruhrmann 2016).

While much research in science journalism has focused on how scientific uncertainty is communicated through text, there is comparatively little research in the use of visual design to communicate about such matters. E.g., in a systematic review about framing research in health communication, Guenther et al. (Guenther, Gaertner, and Zeitz 2021) find a lack of research on visual framing. Some studies in science communication have explored the effect of presentation formats that appear scientific – e.g. using scientific language and charts – on participants' belief in a message, although such claims have been contested (Tal and Wansink 2016; Dragicevic and Jansen 2018; Haard, Slater, and Long 2004; Pandey et al. 2014). A study by Dick (2014) of the use of interactive infographics in news organisations show that organisations often rely on templates, in order to reduce risk.

Treise and Weigold suggest that one of the benefits of science journalism lies in helping the public "better discriminate the activities of scientists from those of "pseudo" scientists" (Treise and Weigold 2002, 311). However, both scientists and journalists express frustration with science journalism, and an obstacle in this effort is the absence of "solid metrics" (Treise and Weigold 2002, 320). The accuracy and rigor of science journalism has been the subject of much controversy, but assessing the quality of science journalism is far from trivial ("Science Journalism Can Be Evidence-Based, Compelling - and Wrong" 2017). Šuljok and Vuković (2013) propose a "Trustworthiness Index" for science news, based on whether the news stories cite their sources, include expert opinion and whether the science is presented superficially or in depth. A more recent study has presented a list of 12 quality indicators for science communication, organized in the three categories "Trustworthiness and scientific rigour", "Presentation and style" and "Connection with the society" (Olesk et al. 2021) - however each of these indicators is described qualitatively and broadly and needs further operationalisation in order to be applied.

A notable past effort to address this challenge with particular regard to medical science is the "Index of Scientific Quality" (A. D. Oxman et al. 1993). However, this model is complex and relies on expert input, making it difficult to apply in popular science news. The more recently developed Informed Health Choices framework (Aronson et al. 2019; A. D. Oxman et al. 2018; M. Oxman et al. 2021) is aimed at helping a broader public make critical assessments of health science news, by presenting a list of key concepts that are used to guide the development of learning resources - however, this model places quite great demands on the individual reader, who is tasked with learning a fairly large number of concepts (49) and applying them in critical assessments of health claims, making this approach difficult to apply in science news journalism.

## Fact-checking

Over the last two decades, fact-checking has emerged as an important genre of contemporary journalism (Graves 2016). The proliferation of "fake news" in connection with recent political

events - such as the UK "Brexit" referendum and the election of Donald Trump in the US - have led to much debate about the role of journalism in identifying misinformation and verifying truth (Tandoc, Jenkins, and Craft 2019; Waisbord 2018). Brandtzæg et al. (2018) study the perception of fact-checking and verification services among journalists and social media users, and find mixed opinions in both groups. They offer design recommendations suggesting that such services should focus on transparency, acknowledge limitations and involve lay users in collaborative fact-checking.

In a meta-analysis exploring the effect of fact-checking on political beliefs, Walter et al. (2020) find mixed evidence suggesting that fact-checking has a positive, but weak effect on the accuracy of political beliefs. Studying the effects of correcting misinformation in different domains, Walter and Murphy (2018) find that efforts to correct misinformation are more effective regarding health information, than political beliefs. Focusing on correction of health misinformation in social media, Walter et al. (2021) find encouraging indications that such corrections are effective; and notably, corrections coming from expert sources are more effective than those coming from non-experts.

The COVID-19 pandemic has brought increased attention to concerns about a surge in health misinformation, sometimes characterized as a "misinfodemic" (Krause et al. 2020) or simply "infodemic" (Bechmann 2020). Van Stekelenburg et al. (2021) explore public beliefs about COVID-19 in the US, testing an intervention instructing participants to check whether health claims reflect scientific consensus. Interestingly, they find a high degree of accuracy in the health beliefs among their sample, contradicting prevalent concerns about the spread of misinformation - and they find no effect of the intervention. Schuetz et al. (2021) study social media users' own fact-checking behaviours, and find that fact-checking helps protect users against COVID-19-related fake news. Austin et al. (2021) study the effect of media literacy on protective behaviours during the pandemic, and find that individuals with higher media literacy were more likely to follow recommendations from health experts - highlighting the importance of media literacy for public health during the pandemic.

## Digital tools for science journalism

There exists a variety of digital tools for searching and navigating in scientific literature.

Many such resources are used by researchers and students - see Klucevsek and Brungard (2020) for an overview. For instance, RetractOMatic is a browser extension which highlights retracted publications in literature searches ("RetractOMatic" 2016). One tool that is relevant also for science journalists is Scholarcy, which offers automated structured summaries of research articles, in order to help readers quickly orient themselves about their content ("Scholarcy" 2021). A recent study used Scholarcy to analyze statements about limitations in COVID-19-related studies in medRxiv (Gooch and Warren-Jones 2020). There are also tools developed specifically with science journalists and communicators in mind. For instance, Science Surveyor is an experimental tool developed to automatically provide an overview over the scientific literature on a selected topic (Kirkpatrick 2015). Sci-Blogger is a tool that analyzes scientific papers and generates suggestions for titles for a blog version of the paper (Vadapalli et al. 2018). August et al. (2020) report on an attempt at using natural language processing to computationally identify writing strategies for science communication, with a view to enable future automated tools to support science communication. Tatalovic discusses similar efforts to automate the work of science journalists, suggesting that there may soon be available "start-to-finish science reporting bots" that automate the entire process of generating science news stories (Tatalovic 2018).

While full automation of science journalism may still be a vision for the future, there are tools in current use that support journalists in part of their work. SciCheck is a feature of the larger fact-checking website FactCheck.org which focuses on debunking false and misleading scientific claims ("SciCheck" 2021). Such work is at the moment limited by human resources. A recent report on efforts to create automated systems for fact-checking (AFC) concluded that fact-checking currently requires human judgment that cannot easily be replaced by an automated system, and that "the real promise of AFC technologies for now lies in tools to assist fact-checkers to identify and investigate claims, and to deliver their conclusions as effectively as possible" (Graves 2018, 1–2). Undoubtedly, for the foreseeable future the work of (human) science journalists will be an essential part of securing the quality of science journalism, and distinguishing trustworthy science from pseudo science. At the current state of technology the role of digital tools may be to support the work of the science journalists, not to replace it.

There is a lack of studies that evaluate the ways in which such digital tools as mentioned above affect science journalism and its readers. Some studies have engaged with other stakeholders, such as scientists and science journalists to investigate how digital tools may support and affect their work. Smith and colleagues (2018) interviewed researchers and media professionals in order to identify pain points and potential design solutions, with proposals including recommender systems to help media professionals gauge the relevance of scientific news releases, and better explanations of scientific methods for lay audiences. Of particular interest for this article is a recent paper by Maiden and colleagues (2020) who report on a participatory design process aimed at developing a bespoke digital support tool for science journalism. They report a number of findings that are relevant for this study. First, their informants - experienced science journalists - indicate, perhaps unsurprisingly, that they are under pressure due to lack of time and resources. Furthermore, while the journalists would access a diverse range of digital information sources when looking for story material, less than half reported reading science journals and magazines. However, those who did use peer-reviewed scientific papers would use the prestige of scientific journals as proxies for judging the quality of the papers. While Maiden et al.'s study reveals interesting insights into the ways a bespoke digital tool might address the needs of science journalists, it does not offer any evaluation of how the tool might affect the readers' perception and understanding of the stories produced using the tool. More generally, for all of the tools mentioned above we have searched for studies that evaluate the impact of these tools on readers' perceptions of science journalism, but have not been able to identify any such studies. Thus it should be noted that this impact remains unknown.

The study presented in this article makes a novel contribution by presenting not only the design of a tool for judging the evidence strength and uncertainty of scientific studies, but also an evaluation of the tool using both qualitative and quantitative data from both test users and the general audience. These data allow us to assess the tool not just in relation to the ambitions and ideals of the scientific community and science journalists, but also in relation to the challenges involved in communicating to a real audience through the format of science news.

# Method

This study follows a Research through Design methodology (Zimmerman et al., 2007), in which research findings emerge from reflections on design practice. Research through Design is a well-established approach within design and HCI, and scholars have debated how such approaches may contribute new knowledge. Some scholars emphasize ways in which practical design work can inform theory (Stolterman and Wiberg 2010; Zimmerman, Stolterman, and Forlizzi 2010). Gaver (2012) emphasizes the role of the design artefacts, and suggests that practice-based research should not aim to develop comprehensive theory - but rather that the theory produced by such research should be viewed as "provisional, contingent, and aspirational", and can best be conceived as annotations of realized designs. Other scholars have debated the relation between theory and practice in HCI, giving attention to forms of intermediary knowledge through perspectives such as "strong concepts" (Höök and Löwgren 2012) or "bridging concepts" (Dalsgaard and Dindler 2014).

Research through Design has also been applied in the fields of Media Studies and Journalism, albeit sometimes under different names (Berry and Fagerjord 2017; Løvlie 2011). Nyre et al. use the phrase "medium design" to describe their study in which they report on the design of a location-aware application for journalistic news (Nyre et al. 2012). Doherty (2016) suggests the phrase "journalism through design", arguing for the benefits of Research through Design in journalism research. Experiments with Research through Design in journalism have sometimes been explored in the context of education (Angus and Doherty 2015; Doherty and Worthy 2020; Løvlie 2016; Øie 2013). Research through Design has similarities with action research, which has also been applied in journalism research (Wagemans and Witschge 2019; Hautakangas and Ahva 2018); it has also been discussed in relation to constructive journalism (Løvlie 2018).

In the study at hand we follow both Zimmerman et al. (2007) and Gaver (2012) in highlighting the artifact as a primary contribution, serving as a design exemplar which offers insights for both the research community and for practitioners in journalism and design. The study also has an element of "research in-the-wild" (Chamberlain et al. 2012; Chamberlain and Crabtree 2020), in that the design created in this project has been co-designed in a collaboration with the website's journalists, and implemented in the website and used and evaluated by the website's readers. This offers empirical data about how audiences understand the design and gives ecological validity to our study, avoiding the pitfalls of a purely lab-based experiment.

## The Scientific Evidence Indicator

The Scientific Evidence Indicator (SEI) was developed during 2018-2019 through a collaboration between the Danish science news website Videnskab.dk and researchers at the IT University of Copenhagen, in Denmark. Videnskab.dk is the primary website for popular science news in Denmark, with nearly 500,000 users per month. The website targets a general audience, based on the premise that the website's content should be written to be understandable for anyone who has completed the first year of high school. Commercial audience statistics from Gemius Audience show that the website reaches a broad audience which is quite evenly distributed both regarding gender and age, with males outnumbering females slightly (52% vs 48%) and with a slight overrepresentation of readers younger than 30 years, compared with readers older than 70 years.

The core team for the project consisted of the three authors of this article: the website's head of development (the third author) as well as two university researchers working in the cross-section of design, media studies and journalism (the first and second author). The initial idea for the collaboration came from the website's head of development who wanted to help the readers better understand the strength of evidence and uncertainty presented by the journalists in their articles. Seen from their perspective, the journalists took great care in researching and assessing the scientific studies that were the topics of news stories on the website, in order to ensure that they reported on science in a manner that was both trustworthy and which communicated openly about uncertainty and limitations with the research. However, these considerations were only communicated to readers through the text of the news stories, and not in a structured or formalized way.

The purpose of creating The Scientific Evidence Indicator was to present the data behind the health news in a more structured and understandable manner - both in order to help readers understand the assessment of the individual studies, but also more generally to help readers learn about how to judge the quality and trustworthiness of a scientific study. Thus the goal with this project was not just to create a "quality marker" for news items, but rather to create a learning tool that could raise awareness about scientific standards. In particular, the aim was to help readers recognize and be critical of media reports about studies that do not meet basic scientific standards - i.e. pseudoscience. As an example, at the start of the project the well-known UK newspaper The Independent reported that "Office teabags contain 17 times more germs than a toilet seat" (Barr 2017). This story was further disseminated through news media across the world, however the only documentation of the study at the heart of the story appears to have been a press release from a company selling hygiene products, offering no scientific data to back up the claim. For a casual reader, however, the news stories based on this press release offered little indication that this was not a trustworthy scientific study. Thus from the outset of the project one of the aims was to create a tool which could help readers distinguish such stories - which are not based on trustworthy scientific sources - from stories based on trustworthy science, within the health domain. Furthermore, the website was also interested in how to present different levels of evidence strength and uncertainty among solid scientific studies - such as the difference between single experimental studies testing new hypotheses and large metastudies of the kind health authorities typically rely on when making decisions.

A part of the ambition for the project was that the resulting tool should ultimately be possible to apply not only by the Videnskab.dk website, but also by partnering news organizations - and possibly also by readers themselves, to assess the trustworthiness of scientific studies they might read about elsewhere. For this reason we were interested in exploring whether it was possible to design a tool that could be automated, requiring no input by specialists.

In the following we present the process of designing and evaluating a tool that could address this challenge. We approached this process as a two-part challenge: First, we needed to identify a set of variables that could be used to assess the trustworthiness of a scientific study. Second, we needed to design a prototype that could present this data to readers in a way that was sufficiently simple that it could be understood by the website's target audience, which was defined simply as a general audience with at least the literacy and knowledge of a student in the first year of upper secondary education (typically around 16 years old).

## Identifying variables

The process to identify variables that could serve as indicators of evidence strength and uncertainty was led by the website's head of development along with a team of experienced science journalists. They started out by examining the work practices and methods the journalists would use in their daily work, when assessing the quality of a scientific study within health science. The journalists pointed to peer review, type of method (e.g. as ranked in the Hierarchy of Evidence) and reputation of scientific journals as important indicators for them. This corresponds with the findings in a report by the UK Academy of Medical Sciences on how to label health science news and press releases, pointing to peer review and type of research as the most important indicators of science quality (The Academy of Medical Sciences 2017). Furthermore, the journalists at Videnskab.dk also pointed to the reputation or experience of the individual scientist as an indicator of the epistemic uncertainty. In this case, they relied on a less systematic process where the scientist's title and affiliation but also opinions from other scientists and their own previous experience would weigh in. This corresponds with earlier findings on journalists' decision making processes (Peters and Dunwoody 2016).

The criteria pointed to by the journalists were then mapped against metadata available in the scientific databases or other publicly available sources (e.g. abstracts in PubMed) giving rise to a list of several possible variables. One consideration in the process of selecting these variables was our long-term goal that the SEI indicator might be automated, so that we could develop a software tool to automatically give an assessment of a scientific study. This would make the SEI a versatile tool both for journalists and possibly also for a wider public. This consideration influenced our choice of variables to some degree, biasing us towards measures that could more or less easily be assessed by an automated tool.

The list of variables was refined with input from two experts in information science who were regular contributors to the website. In order to get a further critical expert assessment, we presented the resulting list to a group of experts in evaluating medical science at the Nordic Cochrane Institute. They were critical of the idea of basing an assessment purely on metadata, and suggested that a more qualitative assessment of each study was needed, based on a critical consideration of questions such as: "Do the methods align with 'gold standards' in the research topic", "are the conclusions substantiated by the results" and "do the scientists have conflicts of interest?" This method could not easily be reconciled with our aim to work towards an automated assessment. However, these suggestions led us to introduce a qualitative element in the design, in which the journalists could add remarks based on their attempts at answering the questions asked by Cochrane Institute as well as other notable things they would find in their journalistic research.

This process resulted in the following list of four variables:

*BFI Level*: This variable uses the Danish Bibliometric Research Indicator (BFI) - a national database of peer reviewed scientific publication channels that is used to calculate funding for universities based on their research output - to gauge whether the study has been published in a peer reviewed, meritable scientific publication channel. This variable gives the number 0 if the publication does not meet the BFI's minimum standard, and otherwise 1-3 depending on the BFI score of the publication channel (which depends, among other factors, on the impact factor). In other words, this variable combines two of the indicators suggested by the journalists: Peer review and the reputation of the publication channel. The BFI database is vetted by national panels of experts, and may therefore be considered an authoritative database of well-regarded international scientific publication channels (Danish Agency For Science and Higher Education 2019).

*Evidence Hierarchy*: This variable uses a 7-point model of the evidence hierarchy for medical research, to indicate the strength of the conclusions in the study (see presentation in Fig. 5). This model was based on the models presented by Burns et al. (2011), but were modified somewhat in order to ease comprehension: Some levels were merged, and a separate level for "Animal and laboratory experiments" was introduced between the last and second last level of the hierarchy. The rationale for this addition was that, according to the journalists, it was often hard for readers to understand the difference between in vitro/in vivo research and experiments on animals vs humans. Thus this model represented a version of the evidence hierarchy that had been adapted to privilege ease of understanding for a non-expert audience, on the basis of the journalists' experience in communicating to that audience.

*H-index*: This variable uses the H-index (from SCOPUS) of the highest ranked[1] among the authors of the study as a measure of the "experience" of the scientific team. This variable was the subject of much debate among those involved in the project. The variable was intended to capture the journalists' interest in the reputation or experience of the scientists. While the journalists had suggested that the authors' affiliation could be used as a proxy for reputation, this would both be problematic from an epistemic point of view (e.g., suggesting that the evidence strength of a study depends on the affiliation of its authors would contradict the logic of double blind peer review), and might also be difficult to implement without a universal database of all relevant research institutions. Using the authors' H-index is also highly problematic, as this measure does not say anything directly about the specific study at hand. However, it may be considered a reasonable proxy for the experience of the authors, in the sense that it measures how many studies the author has published that meets a certain quality threshold (measured by citations). As can be seen in Fig. 6 we opted not to use the exact number but rather created a custom scale with just four levels: "Excellent" (60+), "Very Experienced" (40-60), "Experienced" (20-40) and "Less Experienced" (0-20). This scale was established by considering a sample of medical science articles that the website had written about in the past, and the H-indexes of their authors.

*Special Remarks*: Used by journalists to point out important aspects of a study that are relevant when assessing the reliability of the conclusions. This field may be empty, however journalists are *required* to add an explanation if they are reporting on a study that has not been peer reviewed according to the BFI standard.

During the process of developing this model, the university design researchers (the first and second author) had a concern that the model might be too closely shaped by the journalists' work practices - so that it might end up simply as a validation of their existing practice, rather than offering a genuine assessment of scientific evidence. In order to test this assumption, the designers asked the journalists to apply the model to all the scientific studies in health that they reported on for a couple of weeks, and share the results with the designers. Somewhat to our surprise, several of the studies reported on by the journalists scored 0 on the "scientific publication" variable, because they were published in channels not approved by the BFI system. In one case, the study in question was published in a journal that had been included in "Beall's list of potential predatory journals and publishers".[2] Searching through past news items on the website revealed further examples of studies that were not approved in the BFI system. This was an eye-opening finding for both the designers and the journalists.

---

[1] The rationale for using the highest ranked author - instead of e.g. taking the average or median among all the authors - was that since all authors were assumed to have read and approved the study, each author may be considered to give the weight of their authority to the study - such that e.g. a study in which several junior researchers collaborate with one senior researcher should be weighed similarly to a single-author study by the same senior researcher.

## Designing the prototype

From the outset it was envisioned that the SEI would work as a module that would be attached to news items about medical research. In parallel with the process identifying the variables to be used in the module, a series of prototypes were designed in order to explore how to present these data to viewers. Throughout the design process a central concern was finding the right balance between simplifying the presentation and acknowledging the complexity (and limitations) of the assessment of the scientific studies. Assessing the evidence strength of a scientific study involves a number of careful considerations and questions which do not always have fully general answers backed by clear scholarly consensus. However, from the outset we assumed that readers might only be willing to spend a very short amount of time looking at the SEI indicator, and testing early sketches and prototypes on university and high school students confirmed this assumption. Thus communicating an assessment that was both complex and contested in this format was a considerable challenge that stayed at the heart of the design process throughout the duration of the project. In the following we will explain how we addressed this challenge on four separate levels: Presenting the variables, giving an overall assessment, shaping the presentation, and designing coherence in the presentation.

The first level of the design challenge was presenting and explaining the variables in the model. We saw a need for "translating" the scientific terminology into terms which were understandable to non-specialist readers, while also staying reasonably close to the precise scientific meaning. After several iterations of lo-fi prototyping and testing on students, this led us to rename the three main variables, so that "BFI Level" became "Scientific Publication", "Evidence Hierarchy" became "Research Method", and "H-index" became "The Researcher's Experience". Arguably, these "translations" are somewhat less precise than the original terms, but also slightly easier for non-specialists to understand - a trade-off we had to repeat in many elements of the design, including graphic elements. For instance, we decided to include a graphic representation of the scale for each variable, as can be seen in Figures 4-6. The scale for the H-index (Fig. 6) caused us particular effort, as we searched for clip art that could illustrate a highly experienced and an inexperienced scientist. An early version in which the inexperienced scientist was represented as a baby in a lab coat was rejected for fear of appearing disrespectful towards researchers with a low H-index. We were somewhat unhappy that the final result failed to represent gender diversity - as both figures appear male - but did not have time or resources to make our own graphics.

---

[2] https://beallslist.net/

The next level of the design challenge was helping readers see an overview of the assessment in order to reduce the initial complexity for someone first encountering the SEI. Already in the initial idea for the project, the design had been envisioned as a "credibility barometer" - that is, a "meter" figure that could offer a "reading" of the degree of credibility of a scientific study. In order to realize this idea, we created several alternative models for determining an overall "credibility score" based on the four variables, along with a range of visual representations inspired by meters and gauges. However, these design proposals led to resistance from the website's journalists, who feared that such a meter figure would appear too blunt and simplistic. The journalists worried in particular that this might cause angry reactions from the scientists they relied on as sources, if scientists would see their work presented alongside a meter figure indicating low credibility. For this reason we designed some lo-fi prototypes of the SEI which did not include such a visual gauge, but either just presented the score for each of the four criteria without any aggregated measure, or included a verbal summary of the assessment saying: "The conclusions in this study can be trusted to a low/moderate/high degree." However, testing these designs showed that they were confusing to readers, leading us to conclude that a visual gauge was needed in order to offer readers an overall guide to the assessment. After some deliberation the website's journalists and editors agreed to this choice, deciding that helping readers understand the design took priority above the concern for the scientists' reactions. Thus the SEI was designed to rate to what degree readers could trust the conclusions in a scientific study: To a low, moderate or high degree. This rating was done according to a formalised logic, created by the website's journalists: The "low" category was used for those studies that were placed on the lowest level on any of the two variables Scientific Publication or Research Method - meaning that the study was either not published in a channel that met the BFI system's standards, or used methods placed at the bottom of the evidence hierarchy. For those studies that did not fall into the low category a score was calculated based on the three quantitative variables (not including the qualitative "Special Remarks" variable), and this score determined whether the trustworthiness of the study was rated "moderate" or "high".

The complexity of the textual and visual information included in the SEI model also meant that we needed to spend much work on the shape and placement of the SEI module in relation to the news article. Initially the SEI was envisioned as an "information box" that would present the assessment along all main variables. The size and complexity of this box meant that it needed to be placed at the bottom of the news article. However, early testing demonstrated that this was not viable, as most test users simply didn't read till the end of the news article - and thus never encountered the design unless specifically directed by the designers running the test. Instead we designed a system which divided the presentation in several levels of increasing detail and complexity, and which due to its small size and shape could be placed prominently at the start of the news story, just below the top image and before the article body. The entry level of the design consisted of a relatively small figure that presented a meter figure along with the words "Scientific Evidence Indicator", followed in smaller font by a question: "At a glance: How trustworthy is this research?"[3] (Fig. 1 and 2). For studies that satisfied the basic criterion for peer review (that is, a BFI level of 1 or higher) the figure also included the words "PEER REVIEWED" on a green background, whereas if the study failed this criterion it would say "NOT PEER REVIEWED" on a red background. The needle in the meter figure also indicated the overall assessment of the study, although without labels to explain the assessment. Thus, at this first level the design primarily communicated whether or not the study had passed scientific peer review, without offering much more detail. The second level of the design could be accessed by clicking the figure, which would bring up a larger information box (in UX terms known as "modal box") which presented both an overall assessment and an explanation of the four variables (see Fig. 3-7). While this box gave room for a fair amount of information, we saw a need for a further level where we could present more detailed and more qualitative information about issues of importance for the assessment, caveats, limitations etc. For this purpose we set up a separate webpage which discussed the most important issues with each part of the assessment in some detail, and placed links to these discussions from the relevant parts of the text in the modal box. That webpage also included some more general educational material created by the journalists about how to distinguish between reliable science and unscientific statements or pseudoscience, including a presentation of the website's "manifesto" for critical science journalism.

---

[3] Our translation, with helpful input from Atle Frenvik Sveen. The phrasing "at a glance" was intended to indicate that the SEI assessment should be seen as a heuristic rather than an in-depth assessment.

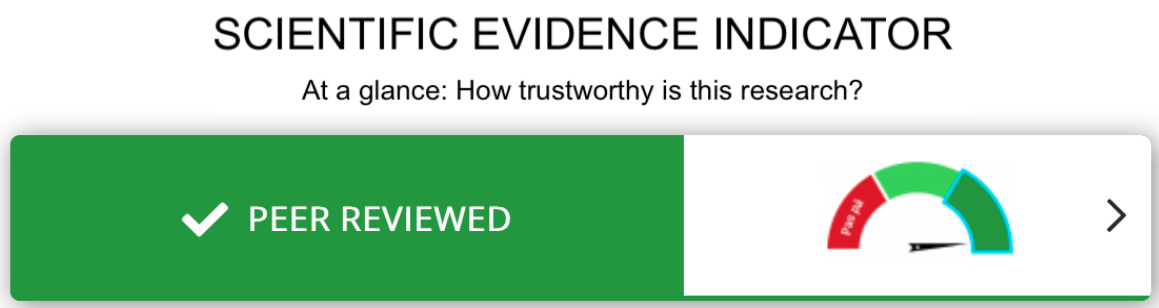

*Figure 1: The entry-level element of the Scientific Evidence Indicator. All the screenshots of the design have been edited to insert English-language translations by the authors.*

# Forskere: Misinformation om HPV-vaccinen kan udløse 45 unødvendige dødsfald

Tusindvis af danske piger valgte HPV-vaccinen fra efter dårlig omtale i medierne. I værste fald betyder det kræfttilfælde og dødsfald, som kunne være undgået, lyder det i nyt studie.

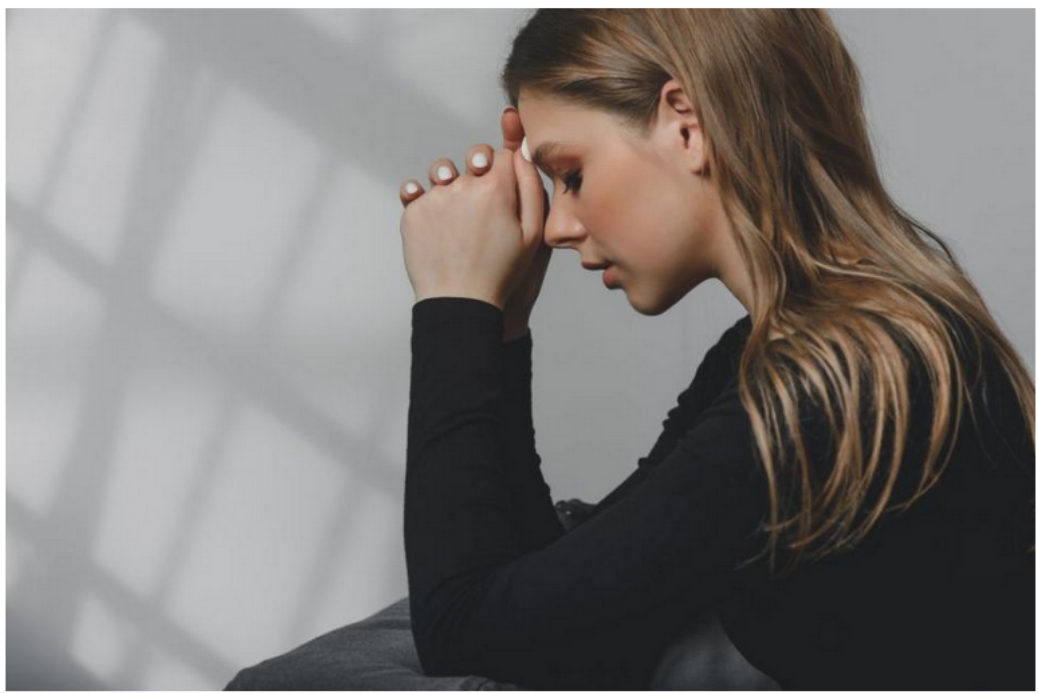

Ingen vaccine er lig med dårligere beskyttelse mod dødbringende virus. Efter en nedgang bliver flere piger dog nu igen vaccineret mod HPV. (Foto: Shutterstock)

**Thomas Hoffmann**
Journalist · Twitter: @Tuffmann

○ 10 januar 2020   (MEDICIN)  (VACCINE)  (SYGDOMME)

### EVIDENSBAROMETER

Hvor meget kan jeg umiddelbart stole på denne forskning?

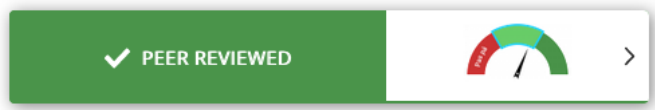

✓ PEER REVIEWED

Det kan vise sig at have store omkostninger, at den danske mediemølle for få år siden kværnede løs med budskabet om, at HPV-vaccinen måske gjorde danske piger syge.

*Figure 2. The entry-level element of the Scientific Evidence Indicator, appearing below the leading image and before the article body.*

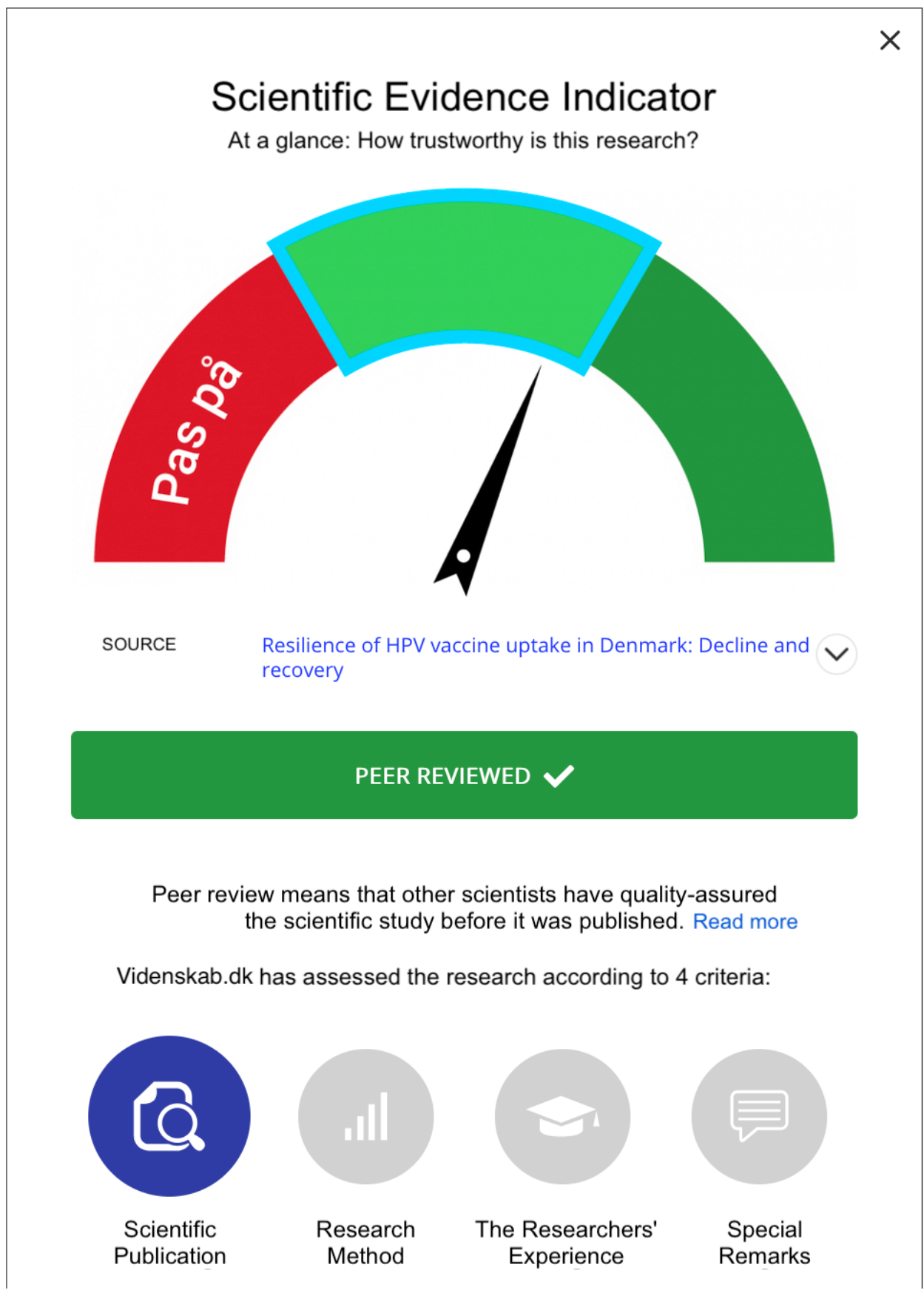

*Figure 3: The top half of the expanded modal box, with the Scientific Publication variable selected. (The label on the red part of the gauge says "Beware" in Danish.)*

Scientific Publication

**BFI Level 3**
Peer reviewed and published by a highly renowned journal

**BFI Level 2**
Peer reviewed and published by a renowned journal

**BFI Level 1**
Peer reviewed and published by an approved journal

**BFI Level 0**
Not peer reviewed and not published by an approved journal

Scientific research is usually published in journals, which are ranked. Only very few get 3 BFI points. Among the most renowned are Science, Nature and New England Journal of Medicine.

A BFI score of 2 is also good. There are many good journals in this category.

A BFI score of 1 is normal. The research has been peer reviewed and is credible.

Solid medical research is always peer reviewed. That means that unbiased scientists have read and approved the content. Journals that do not do peer review are not approved, and their conclusions are uncertain - at best.

The conclusions in this study can be trusted to a moderate degree

Read more >

*Figure 4: The bottom half of the modal box, showing the presentation of the "Scientific Publication" variable.*

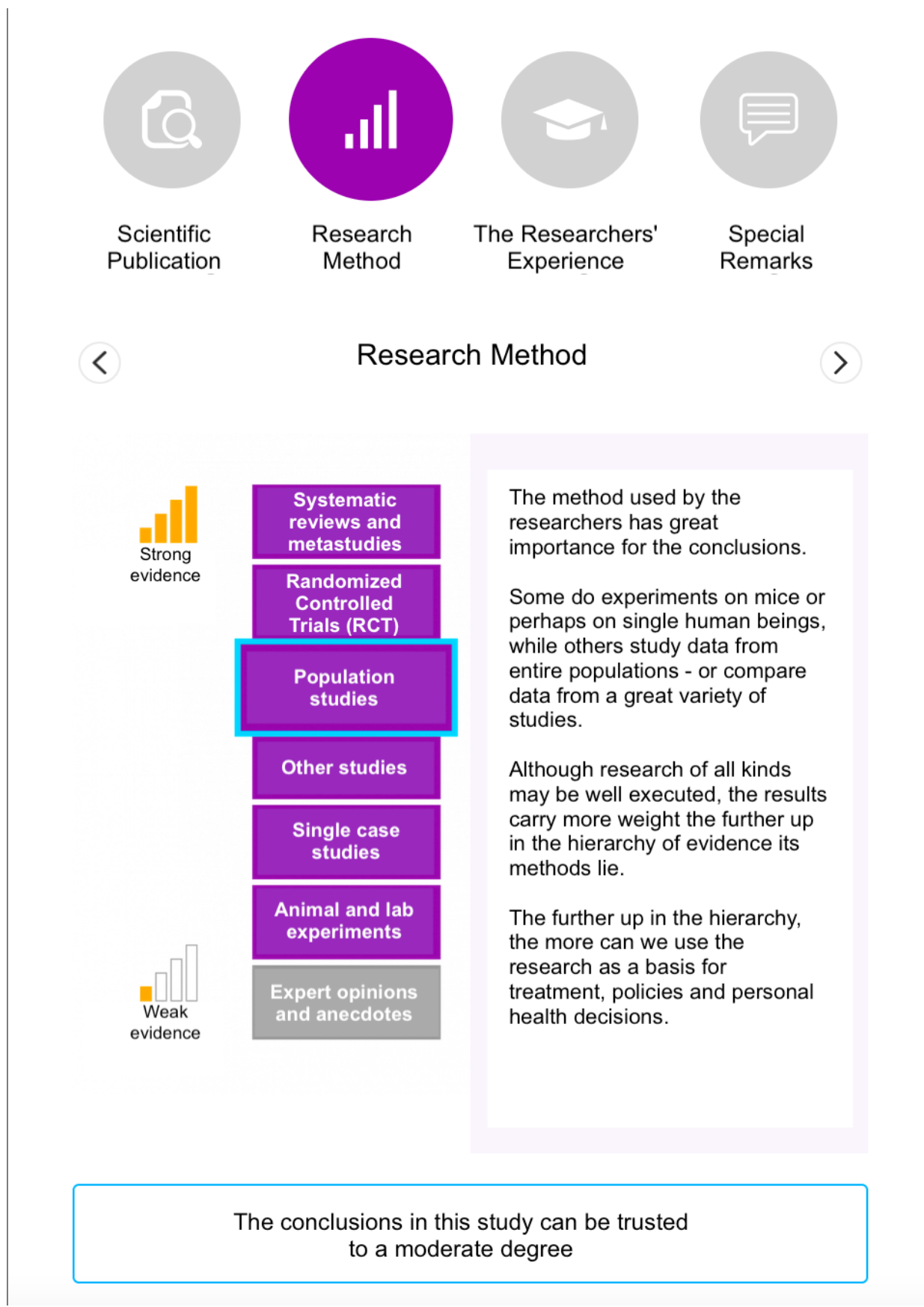

*Figure 5: The presentation of the "Research Method" variable, showing the adapted model of the evidence hierarchy.*

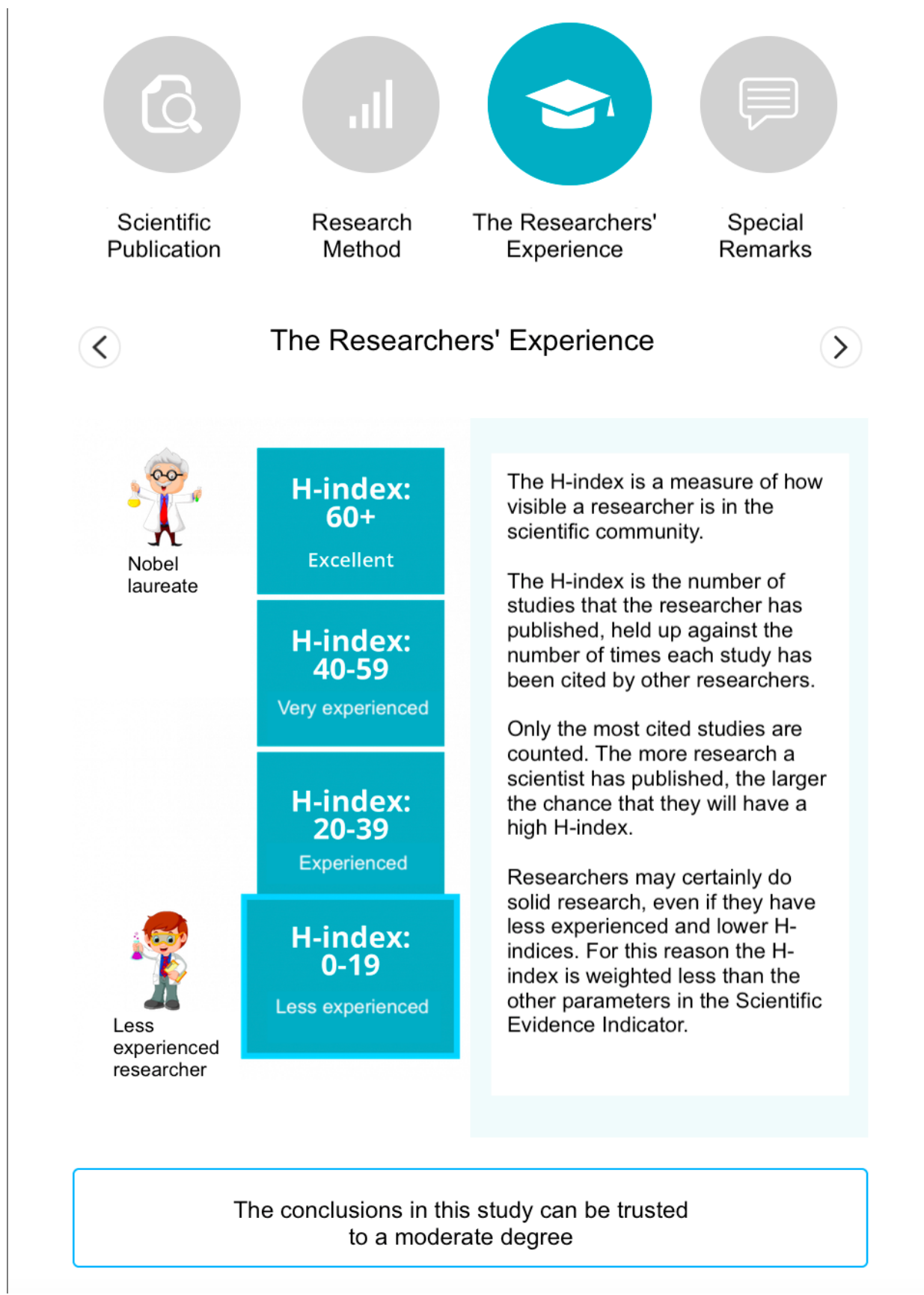

*Figure 6: The presentation of the "Researcher's Experience" variable.*

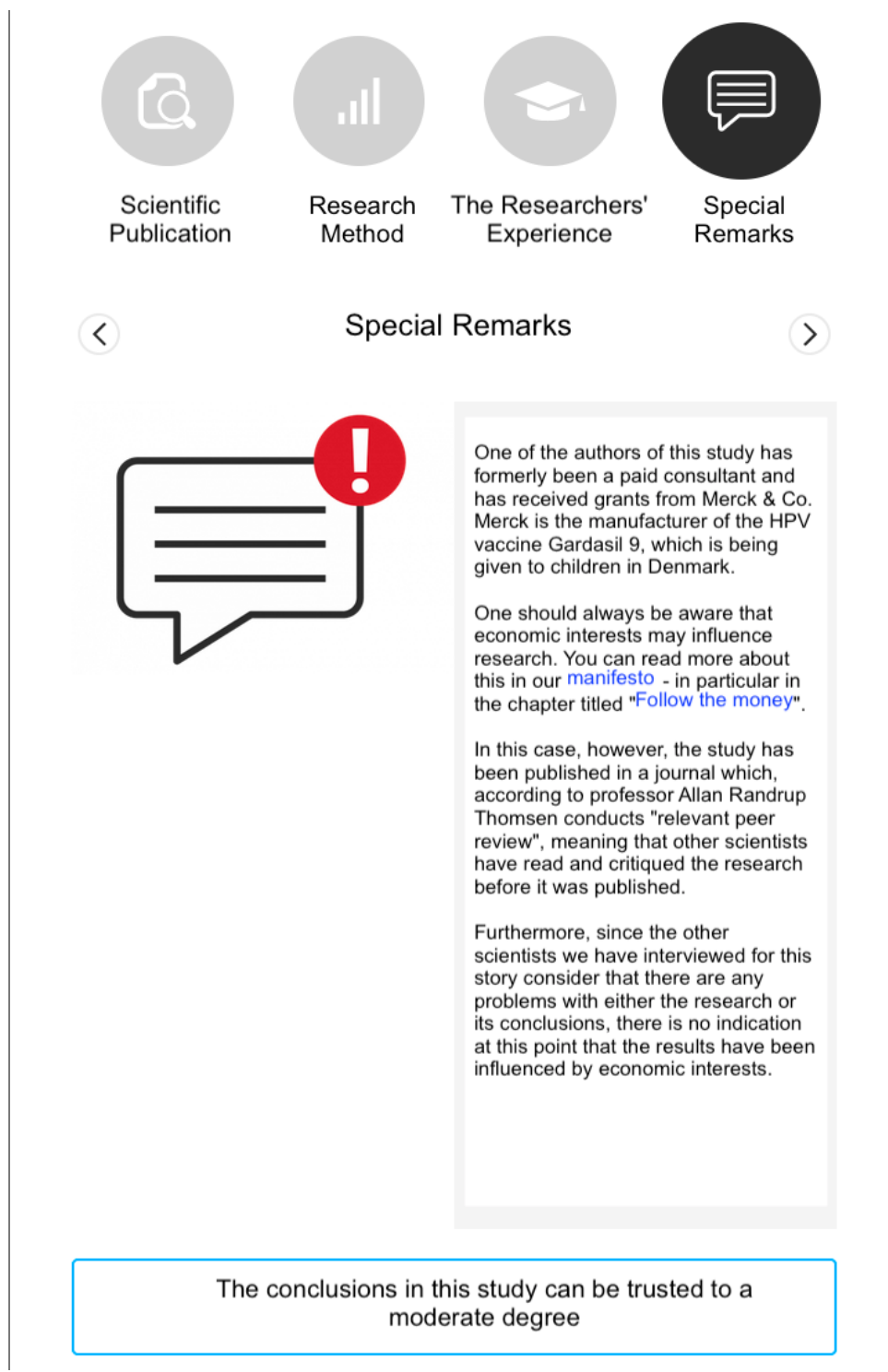

*Figure 7: The presentation of the "Special Remarks" variable.*

A final challenge for the design was creating a visual and logical coherence between these three information levels, to guide the readers and help them understand the relationship between the different elements. This required designing a logical progression from the entry level of the SEI (the initial "anchor" figure, Fig. 1) to the first (top) part of the expanded modal box (Fig. 3), which repeated the main elements of the anchor: The title, the question, the gauge (now with labels) and the text box indicating peer review. Below this level followed a presentation of the four variables, represented by four symbols that could be clicked to access a presentation of that variable (Fig. 4-7). Each variable was presented with a scale that included both verbal labels for each step in the scale as well as a small graphic illustrating the end points of the scale. Clicking on a label in the scale would produce a short explanation in the text box to the right. Below the presentation of the variables followed a text box which presented verbally the overall assessment, saying with words the same that was communicated by the initial gauge: "The conclusions in this study can be trusted to a low/moderate/high degree". The final element was a link saying "read more", leading readers to the separate webpage that constituted the third level of the information design.

*Prototype test*

Towards the end of the design process, before handing the design over to technical developers for implementation in the website's content management system, we conducted a test of a high fidelity clickable prototype to assess whether the design was comprehensible to readers in the target audience. This test was carried out in October 2018 in a high school class consisting of 14 students aged 16-17, in the context of the classroom. The students were presented with three recent stories from the website about the health effects of cannabis, which had been incorporated in a prototyping tool in which an interactive prototype for the SEI had been implemented. The three articles used scientific sources which were assessed differently by the SEI prototype: One was rated as having low trustworthiness, another as moderate and the last as high. After reading the three articles, the students were asked to answer an online questionnaire and participate in a short group interview. Nearly all the students were able to correctly state the "evidence level" of each study (according to the SEI scale, assigning a "low/medium/high" level to each study) in the online questionnaire. As the students were allowed to look at the articles while filling out the questionnaire, this fact simply verifies that the students were able to locate and understand the overall assessment from the prototype. However, only around one third of the students were able to correctly identify that the assessment was based on the four variables presented above, and when asked to explain in their own words the meaning of these variables, most struggled to give a clear answer. This seemed to indicate that the SEI design was effective at *informing* the students about the scientific evidence level for each of the three articles. However - perhaps unsurprisingly - our design did not succeed in giving the students a more fundamental *understanding* of the variables used in the model and what they reveal about the scientific evidence strength and uncertainty for each study - at least not in this limited test setup. The test led us to make several adjustments to simplify the design and ease comprehension.

## Evaluation

The SEI was implemented on the Videnskab.dk website in August 2019, as an element in the website's content management system that was added to news stories about research in the medical sciences. The SEI is still in use at the time of writing this article, in June 2021.

To evaluate the final design an A/B test was conducted on the website six months after the SEI had been implemented. For this test, we used 5 news stories on the website about new scientific discoveries relating to health. The topics covered in the articles were either novel insights or somewhat controversial studies adding new findings to ongoing discussions about the issues: HPV vaccines, air pollution and antipsychotics (see Table 1). Some of the articles were surprising and not in complete accordance with either previous research or the public opinion: "Scientists: Misinformation on HPV Vaccines May Cause 45 Unnecessary Deaths", "Children from Areas with High Air Pollution Can Develop Schizophrenia When They Grow Up", "Many Side Effects from Antipsychotics and Very Little Evidence". One article was more of an opinion and case borne piece but still based on new findings: "Scientists Recommend: Choose Antipsychotic Medicine Based on its Side Effects". Three articles were rated 'high' by the SEI (two of these shared the same primary source), and two were rated 'moderate'. In all five articles the journalists interviewed scientists not involved in the studies to comment on the quality of the study. In four articles the independent scientists confirmed the value of the findings; in the last case the independent scientist wanted more data before concluding ("Many Side Effects from Antipsychotics and Very Little Evidence").

*Table 1: News items used in the A/B test.*

| | News item | Date | Primary source | URL | SEI score |
|---|---|---|---|---|---|
| 1 | "Scientists Recommend: Choose Antipsychotic Medicine Based on its Side Effects" | 9 December 2019 | Huhn et al. (2019) | https://videnskab.dk/krop-sundhed/forskere-anbefaler-vaelg-antipsykotisk-medicin-ud-fra-bivirkninger | High |
| 2 | "Many Side Effects from Antipsychotics and Very Little Evidence" | 18 December 2019 | Huhn et al. (2019) | https://videnskab.dk/krop-sundhed/antipsykotisk-medicin-har-store-bivirkninger-og-sparsom-evidens | High |
| 3 | "Children from Areas with High Air Pollution Can Develop Schizophrenia When They Grow Up" | 7 January 2020 | Horsdal et al. (2019) | https://videnskab.dk/krop-sundhed/boern-fra-omraader-med-meget-luftforurening-udvikler-oftere-skizofreni-som-voksen | Moderate |
| 4 | "Mysterious Symptoms Affect Teenagers - whether They Have Had the HPV Vaccine, or Not" | 9 January 2020 | Thomsen et al. (2020) | https://videnskab.dk/krop-sundhed/gaadefulde-symptomer-rammer-teenagere-uanset-om-de-har-faaet-hpv-vaccine-eller-ej | High |
| 5 | "Scientists: Misinformation on HPV Vaccines May Cause 45 Unnecessary Deaths" | 10 January 2020 | Hansen et al. (2020) | https://videnskab.dk/krop-sundhed/forskere-misinformation-om-hpv-vaccinen-kan-udloese-45-unoedvendige-doedsfald | Moderate |

For the A/B test the website randomly assigned readers to one of two groups, one of which saw the SEI when reading the story, while the other did not. Readers from both groups were invited to answer a multiple choice questionnaire through a "pop-up" survey that appeared after the reader had spent a few minutes on the webpage. The survey questions were intended to gauge the reader's understanding of the factors used in the SEI model. In the following we present an analysis of the answers to these questions (translated from Danish by the authors):

> *Q1: To what degree would you say that one can trust the research that is presented in this article?*
> - *To a high degree*
> - *To a moderate degree*
> - *To a low degree*

   ○ *Don't know*

*Q2: Do you know what peer review is?*
   ○ *A payment system for scientific articles*
   ○ *A system for assessing the quality of scientific articles*
   ○ *A system for judging scientists that have cheated*
   ○ *A mentoring system for youth*

*Q3: Do you think the research that is presented in this article has been peer reviewed?*
   ○ *Yes*
   ○ *No*
   ○ *Don't know*

*Q4: The medical sciences use a "evidence hierarchy" to assess the method used by the researchers. Can you identify what methods were used in this study?*
   ○ *Systematic review and meta analysis*
   ○ *Random Controlled Trial (RCT)*
   ○ *Population study*
   ○ *Other studies*
   ○ *Case study*
   ○ *Animal studies and laboratory experiments*
   ○ *Expert opinions, ideas, anecdotes*
   ○ *Don't know*

These questions had been formulated with the assumption (based on earlier trials) that they represented an increasing degree of difficulty, in order to gauge the effect of the SEI at increasing depths of understanding. The first question was intended to gauge whether readers in the group which were served the articles with the SEI included actually saw the SEI and could register the overall assessment. The second question was arguably fairly simple, asking readers to identify a general explanation of peer review. The third question challenged readers a bit more, in applying that general understanding to the specific study in question, and assess whether or not it had been peer reviewed. And the final question, challenging readers to identify the research methods used in the study, was assumed to represent a more in-depth level of understanding.

859 readers answered the survey in total (483 of which had been exposed to the SEI, and 376 who had seen the article without the SEI). As can be seen in Figures 8 and 9 older readers and readers with higher education were overrepresented among the respondents: 79% of our respondents state that they have a higher education degree, compared to 34% among the general adult population.

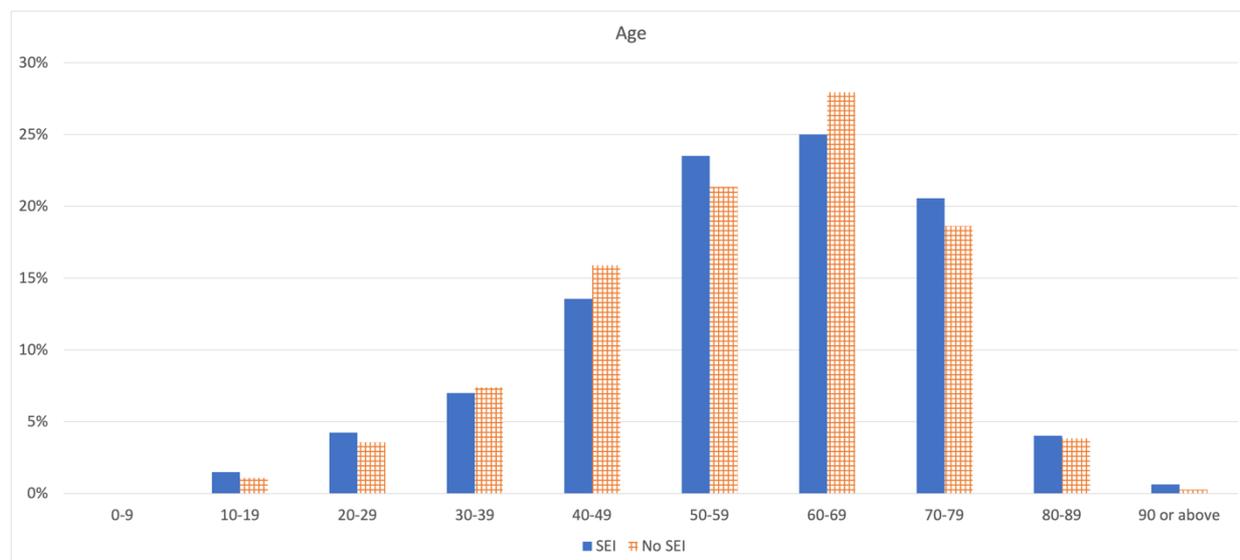

*Figure 8: Age of respondents to the survey.*

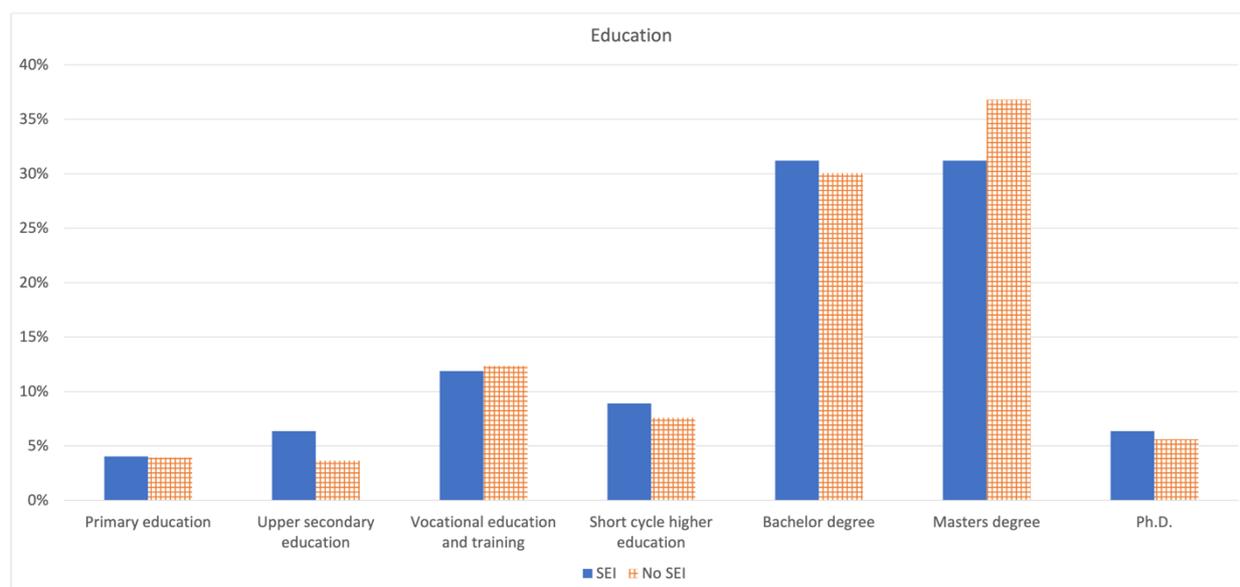

*Figure 9: The respondents' education level.*

Figure 10 shows an overview of the proportion of correct answers to each question for each of the two groups. We test the significance of the difference between the groups through a linear regression in which each question is interpreted as a dichotomous variable (correct or incorrect, blank answers and "don't know" were removed). The difference between groups in Q3 is clearly significant, whereas Q1 and Q2 fall just outside the standard significance threshold. Q4 showed only a small difference between the two groups and is not significant.

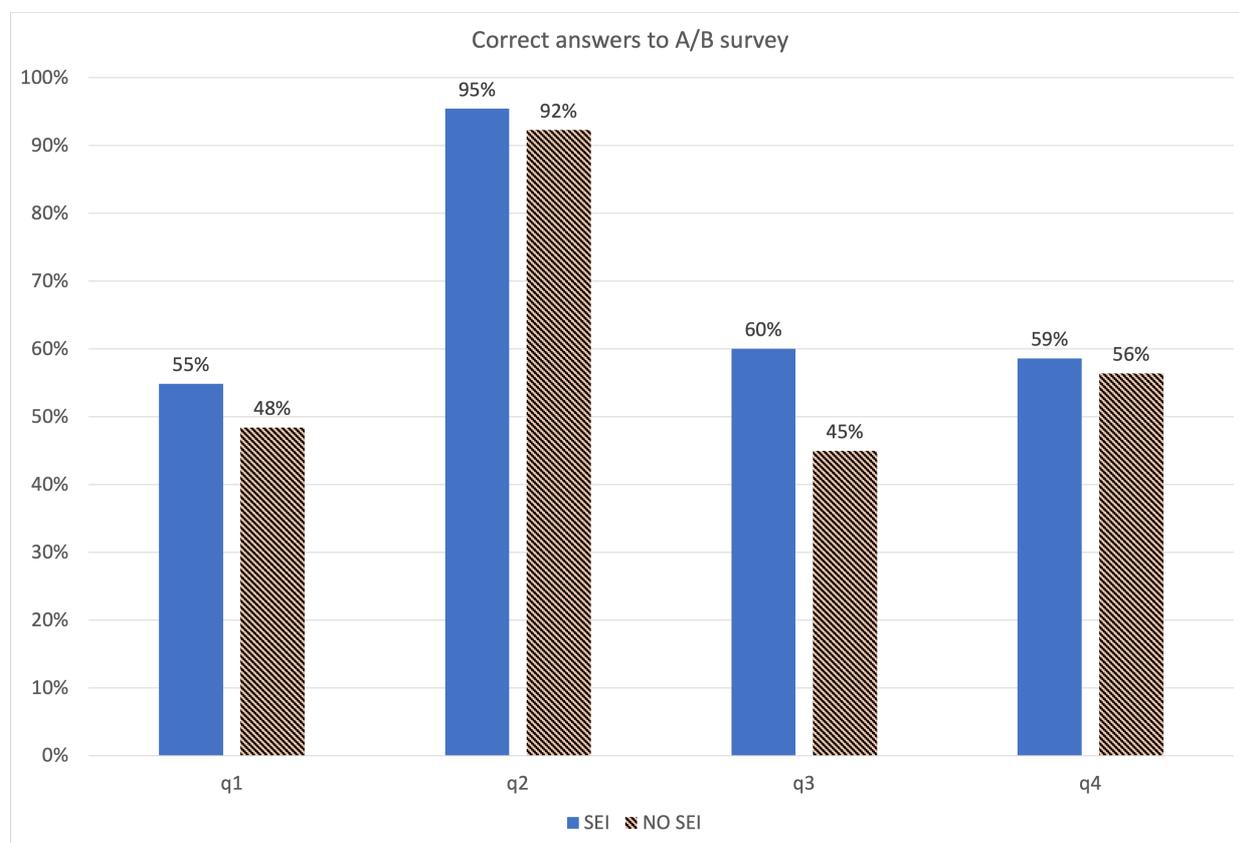

*Figure 10: Correct answers to the survey questions. The columns marked "SEI" represent answers from those readers who were exposed to the Scientific Evidence Indicator, whereas "No SEI" represents the control group. N=859. Significance: p=0.057 for Q1, p=0.071 for Q2, p=0.001 for Q3, whereas Q4 is clearly not significant (p=0.522).*

Note that for the purpose of this analysis we interpret answers to Q1 which align with the SEI output as "correct", although one might argue that this question does not have any clearly defined correct or incorrect answers. For this analysis the purpose is to gauge the difference in the answers between the A and B group, not to determine the epistemological correctness of answers. For Q3 it should be noted that all the five articles included in the test reported on studies that had been published in scientific venues that satisfy the BFI criteria for peer review - meaning that the correct answer to this question should be 'yes' for all five articles.

For all the questions Q1-Q3 there were more correct answers among readers who saw the SEI than among those who did not. While Q3 is clearly significant, for Q1 and Q2 the differences fall just short of the standard significance threshold. These results offer some indication that the SEI had succeeded to some degree in helping readers understand whether the scientific study had been peer reviewed, and whether it was trustworthy (as defined by the SEI assessment). Furthermore, while the vast majority of readers in both groups were able to correctly identify an explanation of peer review, this knowledge is even more present among those who have seen the SEI - indicating that some readers may have attained this knowledge from the SEI. Given the great importance placed on peer review in our initial vision for the design, this result goes some way towards validating that the SEI approach can help readers learn at least some basic knowledge about assessing scientific evidence.

However, the very small difference in Q4 seems to indicate that readers who saw the SEI were not significantly better able to correctly identify the research methods used in the scientific study, than those who did not see the SEI. From our point of view this was somewhat surprising, as this information would be visible to any reader who clicked on the SEI anchor and subsequently on the symbol for the "Research Method" variable (see Fig. 5). A reasonable interpretation would be that few readers did that - in other words, most readers did not spend enough time exploring the SEI to encounter the information that was presented at this depth of the architecture, and which would have helped them answer question Q4. The answers to questions Q1-Q3 were visible either directly on the entry-level SEI graphic (for Q3) or at the top of the expanded modal box (for Q1 and Q2), so readers would encounter this information even after just having had a brief look at the SEI modal box. Thus these results demonstrate both the power and the limitations of the SEI design: The design appears to be effective in conveying an overall assessment of the evidence strength and some fundamental knowledge about peer review, however the design fails to bring across more complex information placed one click further into the architecture.

Figure 11 shows the same data as Figure 10, separated according to which of the five articles the respondents were reading when encountering the survey. Note that most of the respondents came from the first three articles, which show a very similar pattern as in Figure 10. The main variation seems to be in the overall proportion of respondents that have answered correctly on some questions (in particular Q1 and Q3). Presumably this may be due to variation in how the article presents the studies in question. However, the differences between the treatment group and control group remain largely similar for these three articles, suggesting that the impact of the SEI does not depend on the article. The last two articles show some greater variation, but also have much lower numbers of respondents. The very low scores on Q1 for article 5 indicates that respondents tended to agree less with the SEI assessment for this article than for other articles, perhaps reflecting controversies around vaccine research. However, we are reluctant to draw any conclusions from the negative difference between the exposure group and control group for this particular variable as the scores represent a very low number of respondents (5 and 8 respondents, respectively) and may be affected by chance.

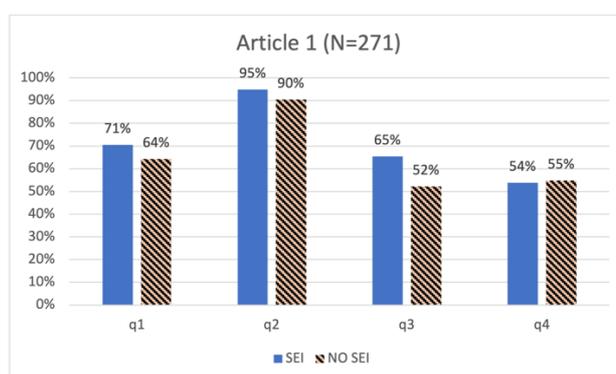

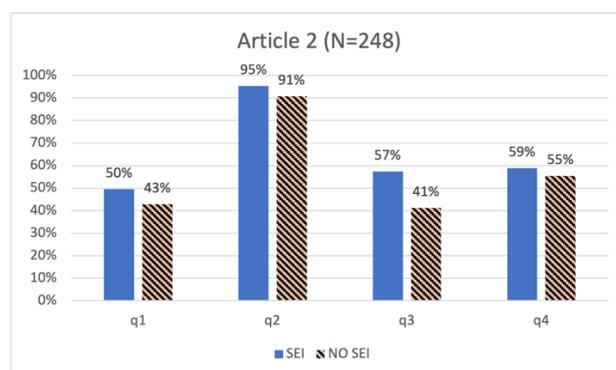

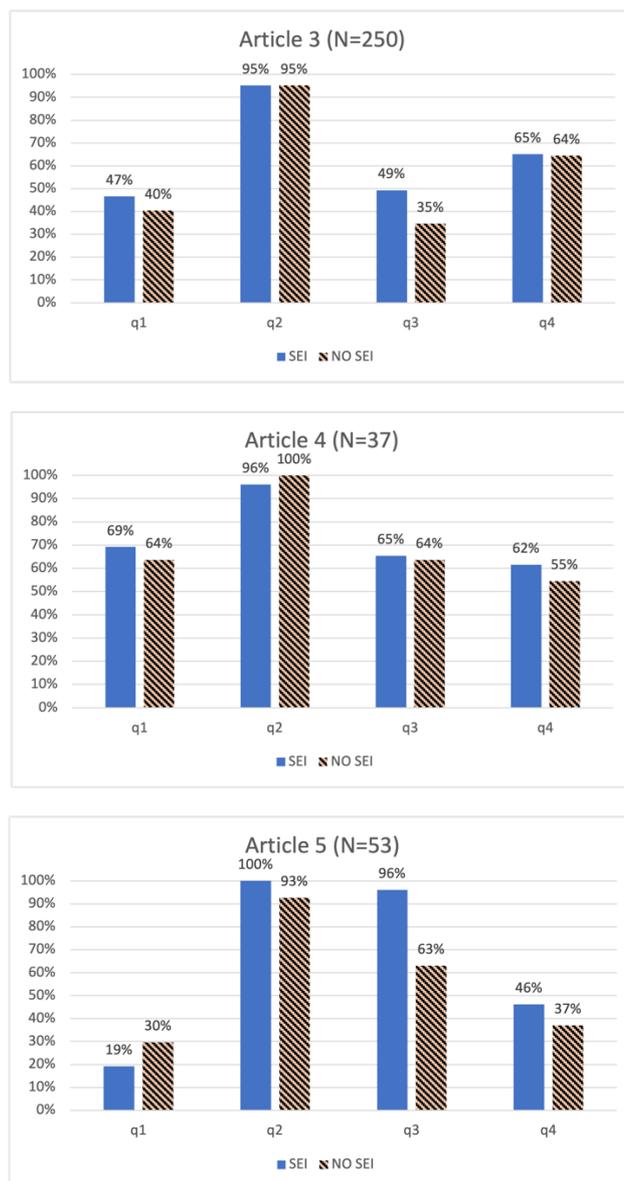

*Figure 11: Correct answers to the survey questions, separated according to the five articles from Table 1. The columns marked "SEI" represent answers from those readers who were exposed to the Scientific Evidence Indicator, whereas "No SEI" represents the control group. N=859.*

*Note that there were few respondents coming to the survey from Articles 4 and 5, and any deviation in the scores reported for these articles compared to the rest of the dataset should be interpreted with caution. For instance, the 100% score on q2 for the control group from Article 4 represents only 11 respondents in the control group from Article 4, whereas the 100% score for the treatment group coming from Article 5 represents 26 respondents in the treatment group from Article 5.*

# Discussion

The results of the A/B test offer some cause for both optimism and concern: The design has shown potential for facilitating learning at a basic level, but there is a challenge if we aim to foster a deeper understanding of how to assess scientific evidence and uncertainty.

The differences in the answers to the A/B survey may seem relatively small, suggesting that the effect of the SEI is somewhat weak. Part of the explanation for this may lie in the fact that the Videnskab.dk website has high standards for the science journalism it presents. Even without the SEI indicator present there are a number of other elements communicating about the trustworthiness of the science - e.g. the articles always cite their scientific sources and include an independent expert assessment in the story, and clearly cite study limitations and other factors that might reduce confidence in the study's results. Since the articles already discuss such issues - and regular readers may thus be relatively well versed in such discussions - the effect of the SEI may be smaller than if we had tested it on a news website with less rigorous standards, or on materials deliberately constructed for the test, drawing on research which did not adhere to scientific standards. The fact that the respondent sample skewed towards higher levels of education may similarly have reduced the effect of exposure to the SEI, compared to an audience with lower education levels. In particular, considering the remarkable high proportion of correct answers to Q2 - about peer review - it may seem that this question was too easy to answer for our respondents to serve as a gauge of the SEI's effect, perhaps due to the respondents' high levels of education.

One might also question whether knowledge about peer review should be treated as a confounding variable. The reason that we treat this question as a dependent variable is that readers in the treatment group (those seeing the article with the SEI module) would be exposed to information about peer review prominently displayed both in the entry-level part of the SEI (Fig. 1 and 2) and in the first view of the expanded module (see Fig. 3, just below the large green box labelled "peer reviewed"). Given that readers were randomly assigned to either the treatment group or control group - and their age and education distributions were similar, as can be seen in Figures 8 and 9 - it is reasonable to attribute any significant difference in the scores on Q2 to the SEI.

The A/B test relies on gauging the effect of the SEI on readers after spending a short amount of time reading one article. Arguably this is a quite shallow exposure, giving readers little time to notice and take in the complex information presented in the indicator. The effect of the SEI might be larger if studied over longer periods of time. Regular readers of the website will encounter the design repeatedly as they read various articles on the site. They might spend time reading the information in the SEI some times, and give it only a casual glance on other occasions - but might gradually learn about the indicators used in the model as they encounter it repeatedly over some time. Since the SEI had been deployed on the website for 4 months when the A/B test started, this factor might also have an effect on the control group in the test: Even though they did not see the SEI on the specific article when they were invited to the survey, they might well have encountered it on articles in the past and thus already have learned about the criteria used in the SEI, leading them to give similar answers to the readers in the treatment group. In order to explore this question it might be illuminating to do a more in-depth evaluation following some readers over a longer time period, for instance in the context of a high school class using the website in educational activities.

Even taking these limitations of the survey into account, it seems clear that the SEI design struggles to convey information other than that which is presented in the initial anchor figure, and perhaps the first view of the expanded modal box (Figure 3). It should perhaps not surprise us if it is hard to entice most readers - even in a website for high quality science news - to turn their attention away from the news article they were reading in order to spend a substantial amount of time and attention on studying complex and intricate details about how to assess the evidence strength of the study. This points to a challenge for future development of the SEI as well as similar endeavors to communicate about scientific evidence and uncertainty: How can one help readers understand more about the complexities and nuances of assessing scientific evidence and uncertainty, without requiring that readers venture far into a complex presentation that takes them away from the main story? One possible path would be to explore ways to convey more in-depth information in the initial graphic that readers encounter, without requiring them to click deeper into the presentation. One might also consider simplifying the presentation in the modal box (Figures 3-7), in order to make it easier for readers to grasp the main insights. Alternatively, one might explore other ways to integrate the information from the SEI into the main news story. One possibility could be to enable journalists to make hyperlinks in the article text to the relevant parts of the SEI module when discussing aspects of the study relating to the criteria presented in the indicator.

However, these challenges also give heightened relevance to a different insight that emerged from the co-design process with the website's journalists: While the SEI initially was intended as a tool for the readers of the news stories, it gradually became clear to us that the design could be just as beneficial - possibly even more so - as a tool used by journalists in researching news stories. As mentioned above, already early in the process journalists had eye-opening experiences when applying the indicator to news stories they were working on. In an internal evaluation conducted by the website's journalists and editors 3 months after the implementation of the SEI, the journalists stated that the work on the SEI project had affected their work practices and made them more aware of issues relating to assessing the credibility, uncertainty and evidence strength of studies that they were considering as topics of news stories. The journalists reported having become more careful in their selection of studies that they based their stories on, and also to ask more pointed questions about uncertainty and limitations relating to the studies when interviewing the authors. Developing the SEI as a working tool for journalists would potentially allow us to design for more in-depth and nuanced assessments, as journalists would presumably be more willing than a casual reader to invest time in studying and understanding the issues presented. This would also shine a new light on the challenge outlined in the paragraphs directly above: The journalists could use the tool in some further depth than readers would, to explore the evidence strength of a study in more detail - and then choose which elements of that information would be most relevant to present back to readers in each case. Possibly the tool could also be redesigned to offer some templates and visualizations that could be useful for journalists when discussing evidence strength and uncertainty of particular studies.

There are a number of limitations with the SEI design. First of all, the model that forms the basis of the indicator is not unproblematic. In particular, using an authors' H-index as an indicator of evidence strength is in logical terms a fallacy, as it does not contain any direct information about the study in question. This aspect of the design was the subject of much discussion between the university researchers - who were skeptical of applying this variable - and the website's participants in the project. The main argument for including this variable was that it represents a heuristic that journalists - and probably also many academic researchers - use when searching for material: If a study has the name of a highly renowned researcher among the authors, most would agree that this study is more likely to be worth taking a closer look at than one whose authors have a less clearly established track record - everything else being equal. Applying this as a heuristic to guide one's attention is not necessarily a fallacy. However, applying it as a measure of the evidence strength of a study is not valid. In future work it should be considered either to remove this variable, or to replace it with other variables more directly indicative of a study's evidence strength.

The other two quantitative variables may seem less problematic. In particular, using peer review as a "gold standard" for trustworthy research is uncontroversial. However, it should be noted that the BFI database which was used for this variable had some limitations - for instance, publication channels that were new or somewhat less known might not be included in the database. Furthermore, any unpublished study - such as a preprint - would by definition receive the score 0 on this variable, as it had not (yet) passed peer review. This means that any study only available as a preprint would be rated by the SEI gauge as having "low" trustworthiness. That may be appropriate, as one should arguably treat research presented in preprints with caution. However, in recent years - and in particular during the ongoing COVID-19 pandemic - preprints have gained increasing importance in science journalism (Fleerackers et al. 2021), and it might be relevant to explore in future work whether it would be possible to give a more nuanced assessment of preprints. In December 2021 the Danish government decided to stop maintaining the BFI indicator, meaning that future versions of the SEI would either need to use another similar indicator - such as the Norwegian Publication Indicator - or find another way to gauge peer review status and the merit of a scientific publication.

The model for creating the overall assessment on the basis of the individual variables is also open to criticism. This model was created by the website's journalists for the purpose of simplifying the overall presentation to readers and does not have a firm basis in any scientific model or theory. Simply put, the model assigns a score for each of the quantitative variables and adds them together. The resulting score determines whether the study is said to be trustworthy to a "moderate" or "high" degree. The "low" category is reserved for studies that score 0 on either the Scientific Publication or Research Method variables - regardless of whether it scores high on other variables - reflecting that a study which is either not peer reviewed or has used a method from the bottom end of the evidence hierarchy should be treated with caution. This model does not represent a scientific method for rating scientific publications, it should perhaps rather be seen as a journalistic simplification. Future work should perhaps explore whether this aspect of the model is communicated clearly to the readers - or perhaps rather, as suggested above, focus on providing the journalists with tools to visualise pertinent data from a more in-depth assessment of the study in question rather than an aggregate of several variables.

Importantly, there are a number of pertinent aspects of scientific evidence and trustworthiness not captured by the SEI design. First of all, questions relating to research funding and conflicts of interest can only be addressed in the "Special Remarks" variable. While the information in this variable does not directly affect the aggregate assessment in the SEI meter, the information is usually also repeated and emphasized by the journalist in the main article. More broadly, as suggested by the group of experts from Cochrane (see above), one might argue that a proper assessment of the evidence strength of a study cannot be done only using metadata about the study, but needs to take the form of a more in-depth assessment of the contents of the study and its appropriate application of theory and methods from the relevant field of research. Furthermore, one might even question the wisdom of assessing evidence strength on the basis of individual studies - and rather favour assessments that hold together evidence from a range of different studies, as is commonly done in systematic reviews. However, while these concerns are valid they might stand in the way of creating a functioning digital tool that can be productively applied in the journalists' daily work.

The SEI also does not capture any aspect of how the journalistic product presents the science - such as whether or not the research findings and any limitations with them are accurately presented or over-hyped, whether headlines, visualizations and other attention-grabbing elements accurately represent the main content of the story, etc. Such issues are subjects of frequent debate, but are beyond the scope of this study. It should be noted that the SEI through its reliance on metadata may privilege research which falls within established paradigms over more innovative and controversial research that challenges the paradigms, as the latter kind of research may take longer to gain acceptance resulting in publication in highly regarded venues, citations et cetera. The SEI indicator is designed to reflect scientific consensus, and may have some conservative bias as a consequence of this. Finally, it should be noted that the results in our study may be contingent on the Videnskab.dk website and its audience, and may not be directly generalizable to other contexts.

## Concluding remarks and future work

In this study, we set out to explore how to design a tool that could help readers understand how to assess evidence strength and uncertainty in scientific studies within health science. The insights gained from the design and evaluation of the Scientific Evidence Indicator has demonstrated both the significant challenges in this endeavor, but also some potentials that are worthy of further exploration. As demonstrated by our final evaluation, the indicator is effective in helping readers identify whether the study they are reading about has been subjected to scientific peer review - a central aim for the project.

We see much potential for future work in designing tools for journalists, both to help them assess the evidence strength and uncertainty of studies they are considering to write about, but also to help them present and visualize such issues for the readers of their news stories. One important concern in such an endeavor would be to find the right balance between providing automated assessments and leaving sufficient room to let the work be guided by the journalists' qualitative, in-depth considerations.

## Funder information

This work was supported by the Google Digital News Innovation Fund.